\documentclass{article}

\usepackage{moreverb}

\usepackage{amssymb}
\usepackage{amsmath}
\usepackage{enumerate}
\usepackage{graphicx}
\usepackage{epstopdf}
\usepackage{latexsym}
\usepackage{url}
\usepackage{multirow}
\usepackage{sidecap}
\usepackage[dvips,colorlinks,bookmarksopen,bookmarksnumbered,citecolor=red,urlcolor=red]{hyperref}

\begin{document}

\title{A Multivariate Cure Model for Left- and Right-Censored Data with Application to Colorectal Cancer Screening Patterns}

\author{\small Yolanda Hagar, Danielle Harvey, Laurel Beckett\footnote{Yolanda Hagar  is a postdoctoral researcher in Applied Mathematics, University of Colorado at Boulder.  Danielle Harvey and Laurel Beckett are faculty in the Department of Public Health Sciences, University of California, Davis. Correspondence emails: {\tt yolanda.hagar@colorado.edu}}}
\date{}
\maketitle

\thispagestyle{empty}
\baselineskip 12pt

\begin{abstract}
We develop a multivariate cure survival model to estimate lifetime patterns of colorectal cancer screening.  Screening data cover long periods of time, with sparse observations for each person.  Some events may occur before the study begins or after the study ends, so the data are both left- and right-censored, and some individuals are never screened (the ``cured" population).  We propose a multivariate parametric cure model that can be used with left- and right-censored data.  Our model allows for the estimation of the time to screening as well as the average number of times individuals will be screened.  We calculate likelihood functions based on the observations for each subject using a distribution that accounts for within-subject correlation, and estimate parameters using Markov Chain Monte Carlo methods.  We apply our methods to the estimation of lifetime colorectal cancer screening behavior in the SEER-Medicare data set.\\


\end{abstract}

\maketitle
\section{Introduction}
Colorectal cancer (CRC) is the third most common cancer in the United States, both in incidence and mortality rates \cite{acs2013}.  Because this cancer is largely asymptomatic, it is important for individuals to be screened regularly.  Not only can screening detect colorectal cancer at earlier stages, but it can also detect pre-cancerous polyps, which can be removed \cite{acs2013}.  The effectiveness of CRC screenings led the United States Preventive Task Force to set screening guidelines in 1996 \cite{usptf1}.  In this work, we focus on colonoscopy screenings, which are currently recommended to be performed once every ten years for average risk individuals starting at age 50.  While more expensive and risky, a colonoscopy is the most thorough form of screening as it can examine the entire colon with few false negatives or false positives, and can remove polyps and even some cancers during an examination \cite{WinawerCRCwork}.  Little is known about lifetime colonoscopy screening behavior, as it is challenging to estimate due to incomplete data.  Many individuals are not observed for the entirety of their eligibility, possibly censoring observations that occur before or after the study period.  Additionally, individuals can have zero to many screenings in a lifetime, and can have a screening immediately upon becoming due or can delay varying lengths of time. 

Previous studies have examined rates of colorectal cancer screening in different populations, including the Medicare population (such as those found in \cite{WinawerCRCwork,SeeffCRCwork, SmithCRCwork,  origjama}).  However, while these studies report that screening and adherence rates are low, the methods used do not account for screening behavior that may have occurred before or after the study observation period.  In addition, these studies do not quantify the average amount of time individuals wait between screenings, or how many screenings individuals receive in a lifetime.  Because of this lack of information, it has been difficult for researchers to confirm optimal screening guidelines.  While screening for colorectal cancer reduces cancer risk \cite{acs2013, usptf1, WinawerCRCwork}, the colonoscopy procedure itself can be risky, requires a trained specialist \cite{colonComplLevine,colonComplDominitz}, and unnecessary screenings put an avoidable financial burden on the Medicare system (for example, see \cite{colonoscCostVijan}).  Without knowledge of lifetime screening patterns, it has been difficult to perform long-term cost-benefit analyses for outcomes in colorectal cancer.  Given the importance of determining lifetime colonoscopy screening behavior, we develop a multivariate survival model that allows for a proportion of subjects to never be screened, and we use our model to estimate patterns in lifetime colonoscopy screening behavior.

\section{SEER-Medicare Data}

\begin{figure}
	\centerline{\includegraphics[width=5in, angle = 90]{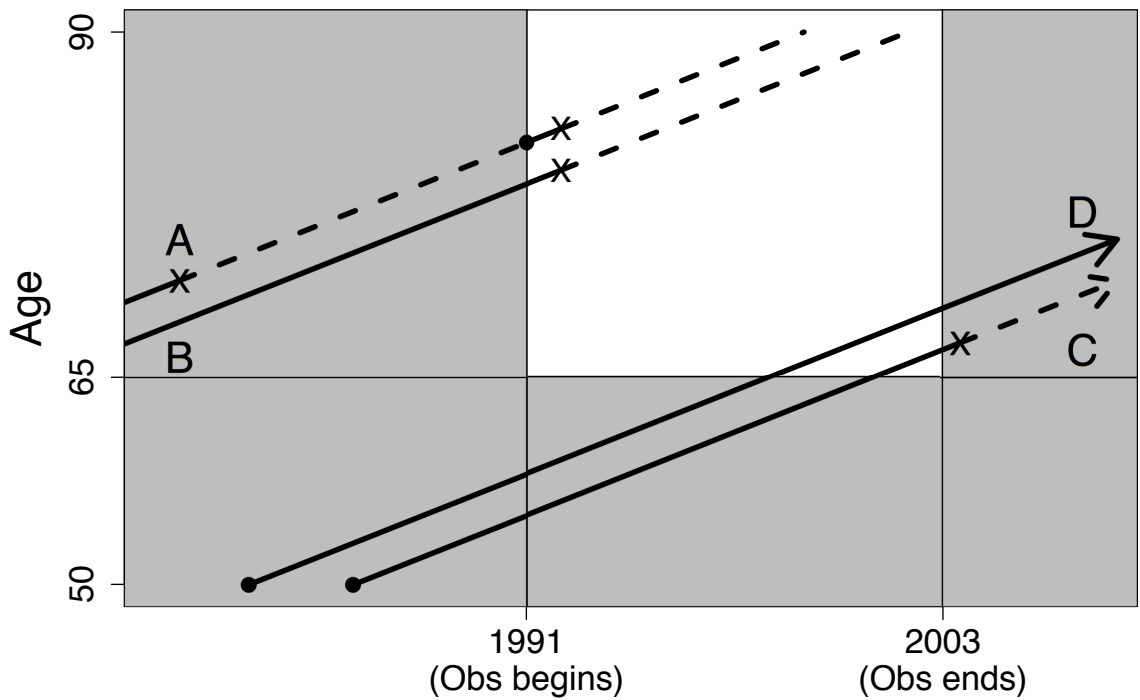}}
	\caption{\footnotesize{Four hypothetical lifetime colonoscopy screening trajectories, and the observed information (unshaded region) available from the SEER-Medicare data set.  In the figure, a circle denotes the time an individual becomes due for a screening (occurs either at age 50 or 10 years after the previous screening), and an `X' denotes the time of a colonoscopy screening.  The solid lines represent periods of time where the subject is overdue for a colonoscopy (called a ``lag time"), and the dashed lines represent the ten year period during which average-risk subjects are not due for a screening.  Among the hypothetical trajectories, subject $A$ is screened twice, once before 1991, and then again in the observation window.  Subject $B$ has only one screening, which is observed.  However, based on information provided in the observation window, it is not possible to tell if the lifetime trajectory for subject $A$ is different than it is for subject $B.$  Similarly, neither subject $C$ or $D$ has an observed screening, but subject $C$ does get a colonoscopy screening after the study ends, while subject $D$ does not.  In this example, all four hypothetical subjects are both left- and right-censored.}}
	\label{fig:SEERMedtraj}
\end{figure}

We used the SEER-Medicare data set to quantify lifetime colonoscopy screening behavior.  This large, public data set is a linkage between the Surveillance, Epidemiology and End Results (SEER) program of cancer registries and Medicare claims files, and is one of the largest and most complete data sets containing colonoscopy screening information \cite{SEERmed}.  However, subjects in the SEER-Medicare data set are age 65 or older, and were only observed between 1991 and 2003.  Possible screenings occurred before age 65, before 1991, or after 2003  and were not observed, so lifetime behavior was left- and/or right-censored for many individuals.  Additionally, some individuals were never screened through colonoscopy, while some individuals were screened more than once in a lifetime.  Examples of the complexities of screening behavior can be observed in Figure \ref{fig:SEERMedtraj}.  Note that although the actual colonoscopy screening behaviors of hypothetical subjects $A$ and $B$ are different, the observed trajectories are the same.  Similarly, hypothetical subjects $C$ and $D$ have identical observed screening patterns but different true lifetime behaviors.  Statistical methods that can account for screening patterns that occur outside the observation window are necessary for proper estimation of lifetime screening behavior.

In addition to estimation of the time and rates of colonoscopy screenings, we are also interested in the impact of health policy changes on screening behavior.  Medicare changed  insurance coverage policy rules in 1998 to provide increased coverage for colonoscopy screenings.  Before 1998, no colonoscopy screenings were covered by Medicare (``phase 0").  Between 1998 and June 30, 2001 (``phase 1"), colonoscopy screenings were covered for high-risk individuals (e.g. those with family history of colorectal cancer), and starting July 1, 2001 (``phase 2"), coverage was provided to all Medicare patients, regardless of risk level.   Previous work has shown that an increase in Medicare colonoscopy coverage led to an increase in screenings \cite{origjama}, however this has not been examined in the context of lifetime screening patterns or in the quantification of the time to being screened.   Understanding the impact of changes in guidelines is important to understanding barriers and patterns of screening behaviors.

The model we propose will answer a number of open questions; we will be able to quantify the number of times individuals are screened in a lifetime, and the length of time individuals wait between screenings, while quantifying how insurance policy changes affect screening behaviors.  To do this, we develop and implement a multivariate cure model that accounts for both left- and right-censoring, within-subject correlation, the estimation of multiple event times, and the average number of events per person.  This model is particularly well-suited for the estimation of lifetime colorectal cancer screening, as these events are sparse over the course of an individual's lifetime and cover long periods of time.  The resulting estimates are robust, despite the left-censoring, right-censoring and number of subjects who are never screened.  The rest of this article is organized as follows:  In Section 3, we discuss the multivariate survival methodology we have developed for left- and right-censored data, including the MCMC algorithm used for parameter estimation.  We also discuss how covariates are incorporated into the model through the parameters.  In Section 4, we present results of a simulation to validate our approach for settings similar to the SEER-Medicare data.  In Section 5, we use our methodology to estimate colonoscopy screening behavior in the Medicare population, using the SEER-Medicare data set, assuming individuals can have up to two screenings in a lifetime.  Section 6 contains a discussion and concluding remarks. 

\section{Model}
Many methods have been used to examine colonoscopy screening behavior (such as those in \cite{WinawerCRCwork,SeeffCRCwork, SmithCRCwork,  origjama}).  These studies examined screening rates using simple approaches, such as counting the number of colonoscopy screenings that occur each year, or more sophisticated approaches using Poisson regression \cite{origjama}.  While these statistical approaches provide basic information on colonoscopy screening trends, to quantify lifetime screening behavior, a model is needed that can account for both the left- and right-censoring inherent in the SEER-Medicare data set, as well as allow for some people to never be screened.  Below, we formulate a multivariate survival model that determines how long people wait between screenings, as well as the average number of screenings individuals get in a lifetime.

\subsection{Background}

Cure models allow for the estimation of time to an event when a subset of the population is risk-free and will never experience the event.  In the estimation of lifetime screening behavior, the event of interest is a colonoscopy screening, and those individuals who will never be screened are part of the population that will never experience the event, and hence are ``cured" in traditional model terminology.  The time to event is calculated as the time an average-risk individual waits between becoming due for a screening (occurs at age 50 or 10 years after the previous colonoscopy screening) and actually getting screened, which we refer to as the ``lag time".   (The lag time is depicted by solid lines in Figure \ref{fig:SEERMedtraj}.)

Cure models have been examined at length.  Initial work by Boag \cite{boag} presents the mixture model $S_{pop}(t; \psi) = \pi + (1-\pi) S(t; \psi),$ where $S(t; \psi)$ represents the survival function for individuals who will experience the event, and $\lim_{t \rightarrow \infty} S_{pop}(t; \psi) = \pi$.  This model has been studied extensively, and can be seen in  \cite{berksGage,haybit59,haybit65,farewellCM,GoldmanCM,MallerCM,sposto92} and others, but can be complex computationally and is difficult to extend to the multivariate case.  

In addition to the possibility that some individuals may never be screened, some individuals may receive many screenings in a lifetime, so the possibility of multiple, dependent events (i.e. colonoscopy screenings) must be accounted for.  To this end, there are many existing multivariate survival models that have been studied.  A large body of work has been devoted to using Cox models and a marginal hazards approach to investigate the effects of covariates on the hazard rate(s), such as that by Wei, et. al. \cite{Wei89}, Liang, et. al \cite{Liang93}, Lin \cite{Lin94}, Prentice and Hsu \cite{Prentice97}, Spiekerman and Lin \cite{Spiekerman98} and others.  These models obtain population-averaged covariate effects, but are mainly attractive when the correlation between observations is not of interest.  Hougaard has done much work with a frailty term in multivariate survival and competing-risks models \cite{HougDistnReview, HougPDF,HougHetero,HougStableDistn,HougFrailty,HougBook}, however, these models do not include the possibility of a cured population.  Extensive work on multivariate survival models has been done by Chen, Ibrahim, and Sinha \cite{ChenIbrPaper,IbrahimBook}, and perhaps matches our work most closely as some of the models allow for a cured population for right-censored data.  In this work, the authors integrate over latent variables representing the number of risks for each subject, as well as a frailty parameter (to account for within-subject correlation) to get a likelihood function that can accommodate multiple events as well as a proportion of subjects who are cured.  However, in addition to right-censoring, we are also interested in the case of left-censoring, and require a model that incorporates this type of missingness. 

We introduce a type of multivariate cure model that allows for the estimation of multiple lag times for each individual and the probability that an individual will receive zero through many lifetime screenings.  Correlation between lag times is accounted for with a frailty term.  

\subsection{Notation} 

Assume an individual has $M$ screenings in his or her lifetime, where $M$ is a random variable such that $0 \le M \le \ell < \infty$, with $\ell$ denoting the maximum number of lifetime screenings possible for any individual.  (The time frame for colorectal cancer screening is finite, so the assumption $\ell < \infty$ is natural.)  The probability an individual will have $m$ lifetime screenings, i.e. $P(M =m)$, is equal to $\theta_m, m = 1,...,\ell$, and the probability an individual will never be screened is equal to $\theta_0 = 1-\sum_{j=1}^\ell \theta_j$.  If individuals are left-censored (i.e. they enter the study after eligibility begins) and/or right-censored (i.e. they leave the study before eligibility ends), we may only observe $k$ of the $M_i = m$ screenings, where $k \le m \le \ell,$ possibly obscuring part of the lifetime screening pattern.  We estimate the lag times and probability of individuals receiving $M$ ($M = 0, \dots, \ell$) lifetime screenings via the likelihood function and the multivariate survival distribution developed by Hougaard \cite{HougPDF}:  

For individuals who receive at least one screening, let $Y_{ij}$ represent the $j^{th}$ lag time for the $i^{th}$ subject ($j = 1, \dots, M_i, i = 1, \dots, n$).  The lag times for each individual are correlated through a subject-specific frailty quantity but within each subject the lag times are independent.  Let $Z_i$ be the subject-specific quantity for the $i^{th}$ subject.  Assume the $Z_i$'s are independent and follow a positive stable distribution with parameter $\alpha$.  The distribution is given by the Laplace transform 
\begin{eqnarray*}
	E\left\{\exp \left(-sZ \right) \right\} = \exp\left(-s^\alpha\right),
\end{eqnarray*}
with $\alpha \in (0,1]$.  The case of $\alpha = 1$ represents the case of independent observations within each subject.  For the $i^{th}$ subject, conditional on $M$ and $z$, the joint distribution for the lag times is then
\begin{align*}
	& P(Y_1 > y_{i1},\dots,Y_M > y_{iM} \mid M, Z = z) = 
		\exp\left\{ -z \left( \Lambda_1(y_{i1})+\dots+ \Lambda_M(y_{iM}) \right)\right\},
\end{align*}
where $\Lambda_j(.)$ is the cumulative hazard of the $j^{th}$ lag time.  For all individuals, the multivariate survival distribution, given $M$, then becomes
\begin{align}
	P(Y_1 > y_1,\dots,Y_M > y_M \mid M) &= 
		\int_z \exp\left\{ -z \left( \Lambda_1(y_1)+\dots+ \Lambda_M(y_M) \right)\right\} p(z) dz \nonumber \\
			& = \exp \left\{- \left( \Lambda_1(y_1)+\dots+\Lambda_M(y_M)\right)^\alpha\right\} \label{eqn:survfxn}\\
			& = S(y_1, \dots, y_M \mid M) \nonumber
\end{align}

Using this multivariate survival function, we can calculate the probability of a colonoscopy screening occurring before, at, or after certain time points.  For example, the probability of a colonoscopy screening occurring at time $y_1$ is calculated as 
\begin{align}
P(Y_1 = y_1, Y_2 > y_2, \dots, Y_M > y_M \mid M) & = -\frac{\partial}{\partial y_1} S(y_1, \dots, y_M \mid M),
\end{align}
and the probability of observing all $M$ screenings can be calculated as
\begin{align}
P(Y_1 = y_1, Y_2 = y_2, \dots, Y_M = y_M \mid M) & = (-1)^M\frac{\partial^M}{\partial y_1\dots\partial y_M}S(y_1, \dots, y_M \mid M)
\end{align}
These calculations are akin to finding the cdf $f(y) = -d/dy S(y)$ in the univariate case.  In our notation, we use $P(Y = y)$ to represent $f(y)$.  While this definition does not mathematically exist for continuous functions, in the multivariate case it is a notational method for expressing an observed colonoscopy screening at time $y$. 

\subsection{Likelihood Function} 

Because the observed screening colonoscopies are sparse, we use Markov Chain Monte Carlo (MCMC) sampling for parameter estimation.  The posterior distribution for each parameter is calculated using the likelihood function, which is formulated below.

Denote the observed data for individual $i$ by $\vec W_i = (t_{Li}, t_{Ri}, t_1,\dots,t_{k_i}),$ where $t_{Li}$ and $t_{Ri}$ represent the left- and right-censoring times (with $t_{Li}$ = 0 if an individual is not left-censored, and $t_{Ri}$ equal to the end of the observation period if the individual is not right-censored), and $t_1,\dots,t_{k_i}$ denoting the $k_i$ observed screening times, with $0 \le k_i \le m_i \le \ell$.  The case of $k_i$ equal to zero denotes that no colonoscopy screenings were observed in the study period.
Using the multivariate survival distribution in equation (\ref{eqn:survfxn}), we can then write a complete data likelihood function as follows:
\begin{align} 
	L(\mathbf{W}, \vec\eta  \mid \boldsymbol\phi, \vec\theta, \alpha) 
		& = \prod_{i=1}^n \prod_{j=0}^\ell \Bigg(P(M = j) p_{ij}(\vec W_i \mid \vec\phi_j, \alpha, M = j)  \Bigg)^{\eta_{ij}}\nonumber\\
		&= \prod_{i=1}^n \prod_{j=0}^\ell \Bigg(\theta_j p_{ij}(\vec W_i \mid \vec\phi_j, \alpha, M = j)  \Bigg)^{\eta_{ij}}, \label{eqn:multLik}
\end{align}
where $\vec\phi_j$ denotes the parameter vector for the survival distribution for $M = j$ lag times, and $\eta_{ij} = I_i(M = j)$ is an indicator variable that equals one if subject $i$ gets $j$ colonoscopies in a lifetime (with $\eta_{i0} = 1-\sum_{j=1}^\ell \eta_{ij}$).  The $p_{ij}(\vec W_i \mid \vec\phi_j, \alpha, M = j)$ are probabilities associated with the $i^{th}$ person having $j$ lifetime screenings, and are calculated using the multivariate survival function in equation (\ref{eqn:survfxn}) based on the screening pattern observed for the $i^{th}$ individual.  An example of the calculation of these probabilities is shown in Sections \ref{sec:univarEx} and \ref{sec:exampleSubj}.  For the case of no observed screenings (i.e. $M = 0$), the probability of zero screenings, $p_{i0}(.)$, is not defined and the only likelihood contribution in this instance is $\theta_0$.  

For subjects who are left- and/or right-censored and who do not have $\ell$ observed screenings, some or all of $\vec \eta_i = (\eta_{i0}, \dots, \eta_{i\ell})$ may be unobserved.  For the unobserved $\eta_{ij}$, the indicators are replaced with their expected values, which are calculated at each iteration of the MCMC routine using the current sampled parameter values. An example of this calculation is shown in Sections  \ref{sec:univarEx} and \ref{sec:exampleSubj}.

\subsection{Example: Univariate likelihood}\label{sec:univarEx}
To illustrate our model in its simplest form, we first cover the univariate case of $\ell = 1$ (i.e. individuals can only get one colonoscopy in a lifetime).  In the univariate instance, the likelihood function in (\ref{eqn:multLik}) can be simplified to:
\begin{align} 
	L(\mathbf W, \vec\eta  \mid \vec\phi, \theta) &= \prod_{i=1}^n \prod_{j=0}^1 \Bigg(\theta_j
			p_{ij}(\vec W_i \mid \vec\phi_j, M = j)  \Bigg)^{\eta_{ij}} \nonumber\\
		& =  \prod_{i=1}^n(1-\theta)^{1-\eta_{i}}\left(\theta p_{i1}(\vec W_i \mid \vec\phi_1, M = 1) \right)^{\eta_{i}}\label{eqn:univLik}.
\end{align}
In the univariate model, subject $i$ who has an observed screening at time $t_{i1}$ contributes the probability $\theta\times p_{i1}(\vec W_i \mid \vec\phi_1, M = 1) = \theta \times f(t_{i1} \mid \phi_1)$ to the likelihood function (as $\eta_i$ is known and equal to 1).  Conversely, subjects who have no observed screenings and who are not left- or right-censored contribute $(1-\theta)$ to the likelihood function, with a known $\eta_i = 0$. 
However, a left-censored subject with no observed screening, who enters the study at time $t_{iL}$, does not have complete information, and it is not known whether a screening occurred before time $t_{iL}$ or did not occur at all.  On this occasion, $p_{i1}(\vec W_i \mid \vec\phi_1, M = 1)=F(t_{iL} \mid \vec\phi_1)$, accounting for a possible screening before study entry.  The value of $\eta_{i}$ is unknown and is updated at each iteration of the Gibbs sampler with its expected value 
$$E\left(\eta_{i}\right) = \frac{\theta F(t_{iL}\mid \vec\phi_1)}{(1-\theta)+\theta F(t_{iL}\mid \vec \phi_1)}.$$  
This expectation is a number between 0 and 1, and therefore a weight is assigned to the probability subject $i$ was screened once or never screened based on when subject $i$ entered the study.  Similar calculations can be made for right-censored or left- and right-censored subjects. Note that in the univariate likelihood, the within-subject correlation parameter $\alpha$ is not necessary, as each subject only has one screening. 

The univariate likelihood is similar to the complete data likelihood presented in Sy and Taylor \cite{sypaper} for the univariate case of right-censored subjects.  This formulation of the cure model does not reduce to the standard cure model as $t \rightarrow \infty$, however this definition allows for more flexibility in the estimation of left-censored screening behavior.  

\subsection{Example Subject} \label{sec:exampleSubj}

\begin{figure}
	\centerline{\includegraphics[width=5in, angle = 90]{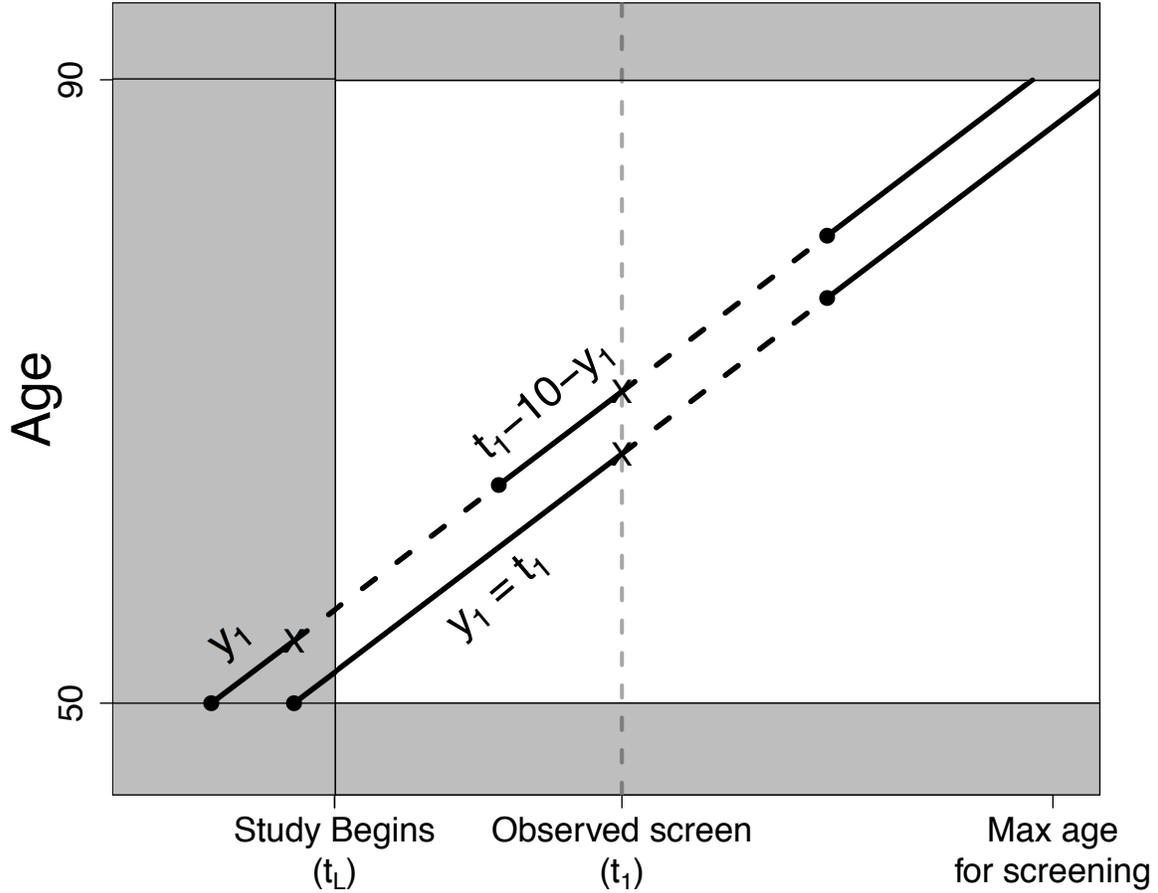}}
	\caption{\footnotesize{The figure shows two possible trajectories for a left-censored subject who enters the the observation period at time $t_L$, and has one observed screening at time $t_1$.  Lag times are denoted with solid lines, and represent periods of time when the subject is due for a screening,  and the ten-year post-screening period is denoted with a dashed line.  Screenings are marked with `X', and time points when the subject becomes due for a screening are marked with a circle.  In this example, it is possible that (1) the one observed screening is the only lifetime screening, or (2) the subject may have had two lifetime screenings, the first one occurring before the left-censoring time $t_L$, and the second being observed.  In the first case, the length of the first lag time, $y_1$, is equal to the time to the first screening, $t_1$, and can be written as $P(Y_1 = t_{1} \mid \vec\phi_1,  M = 1)$.  In the second case, we can only determine that the time to the first screening, $y_1,$ is less than the left-censoring time, and the second lag time, $y_2$ is the remaining period of time between the observed screening and the previous screening ($t_1-10-y_1$).  This can be written as $P(Y_1 < t_L, Y_2 = t_1-10-y_1 \mid \vec\phi_2, \alpha, M = 2)$.  This subject is not right-censored because he or she reaches the maximum screening age before the observation period ends.}}
	\label{fig:exampletraj}
\end{figure}

We now extend the univariate example to the multivariate case of $\ell = 2$.  To prevent identifiability issues, and because the observed screenings are sparse, the application of the multivariate model typically requires the following assumption:
\begin{itemize}
	\item A maximum recommended screening age exists. Since colonoscopy is not risk-free and benefits are long-term (due to the slow progression of colorectal cancer) \cite{WinawerCRCwork}, colonoscopy screening is generally not recommended for average-risk older patients \cite{usptf1}.
	\item The maximum number of lifetime screening colonoscopies, $\ell$ is fixed.  In this example, we set $\ell=2$.  (This is the maximum number observed in the SEER-Medicare data set for average-risk patients).
\end{itemize}

Using these assumptions, we examine the following subject: 
An individual $i$ is left-censored at time $t_{iL} $, is observed until the maximum screening age (i.e. is not right-censored), and has one screening observed at time $t_{i1}$.  Because the individual is left-censored and less than two screenings were observed, there are two possible true trajectories for this subject (see Figure \ref{fig:exampletraj}):
\begin{enumerate}
	\item The one observed screening is the only lifetime screening.
	\item A screening also occurred before observation of the individual began, and so the subject was screened twice in his or her lifetime.
\end{enumerate}
Because there are two possible trajectories, this subject provides a weighted contribution to both the probability that only one lifetime screening occurred and also to the probability that two lifetime screenings occurred to the likelihood function.  The probabilities for both cases are calculated using the distribution in equation (\ref{eqn:survfxn}).  The probability for case 1 (represented by trajectory 1 in Figure \ref{fig:exampletraj}) is calculated under the assumption that the observed screening is the only screening and is written as: 
\begin{align}
	p_{i1} (\vec W_i \mid \vec \phi_1, M = 1) & = P(Y_1 = t_{i1} \mid \vec\phi_1, M = 1)\nonumber\\
		& = -\frac{\partial}{\partial y_1} S(y_1\mid \vec\phi_1, M = 1) \Big|_{y_1 = t_{i1}}.\label{eqn:Meq1ex}
\end{align}	
The probability for the second case (represented by trajectory 2 in Figure \ref{fig:exampletraj}) is calculated under the assumption that one screening occurred before the left-censoring time, and the second screening is observed (i.e. $M = 2$), and is written as:
\begin{align}
	p_{i2} (\vec W_i \mid \vec\phi_2, \alpha, M = 2)&= P(Y_1 < t_{iL}, Y_2 = t_{i1}-10-y_1 \mid  \vec\phi_2, \alpha, M = 2)\nonumber\\
		& = -\int_0^{t_{iL}} \left(\frac{\partial}{\partial y_1} S(y_1, y_2\mid  \vec\phi_2, \alpha, M = 2)\right)\Big|_{y2 = t_{i1}-10-y_1} dy_1. \label{eqn:Meq2ex}
\end{align}	
	  Note that we do not actually observe the length of $y_1$ (the length of the first lag time), we only know that it is less than the left-censoring time $t_{iL}$.  Similarly, we do not observe $y_2,$ the length of the second lag time, and only know that $y_2$ is equal to the length of time between the first screening at time $y_1$ and the second screening at $t_{i1},$ minus the 10-year waiting period.

Because only one screening is observed,the values of $\eta_{i1}$ and $\eta_{i2}$ are not known (although it is known that $\eta_{i0} = 0$), and are therefore iteratively estimated, replacing the unknown values with expected values using the sampled parameter values at each MCMC chain iteration.  In this example, the expected value at the $r^{th}$ MCMC iteration is calculated as:
\begin{align*}
	\eta_{i1}^{(r)} &= E\eta_{i1}^{(r)}
		= \frac{\theta_1^{(r)} p_{i1}(\vec W_i \mid \vec\phi_1^{(r)}, M = 1)}
			{\theta_1^{(r)} p_{i1}(\vec W_i \mid\vec \phi_1^{(r)}, M = 1) + \theta_2^{(r)} p_{i2} (\vec W_i \mid \vec\phi_2^{(r)}, \alpha^{(r)}, M = 2)}\\
	\eta_{i2}^{(r)} & = E\eta_{i2}^{(r)}
		= \frac{\theta_2^{(r)} p_{i2} (\vec W_i \mid \vec \phi_2^{(r)}, \alpha^{(r)}, M = 2)}
			{\theta_1^{(r)} p_{i1}(\vec W_i \mid\vec \phi_1^{(r)}, M = 1) + \theta_2^{(r)} p_{i2} (\vec W_i \mid \vec\phi_2^{(r)}, \alpha^{(r)}, M = 2)},\\
\end{align*}
where $p_{i1}(.)$ and $p_{i2}(.)$ are calculated using equations (\ref{eqn:Meq1ex}) and (\ref{eqn:Meq2ex}).

\subsection{Parameter Estimation} \label{sec:gibbsest}
Because the likelihood is high-dimensional and the observed screening colonoscopies are sparse, we use Gibbs sampler \cite{Gibbs} to estimate the posterior distributions of each parameter, iterating through the following steps:
\begin{enumerate}
\item \label{item:thetasample}  Draw $\vec\theta$ from the conditional posterior distribution
	\begin{align*}
	\Pi(\vec\theta \mid \vec\gamma, \boldsymbol\eta) &
		\propto \left\{ \prod_{i=1}^n \prod_{j = 0}^\ell \theta_j \right\} \times Dir\left((\gamma_0, \dots, \gamma_\ell)^T\right) 
		= Dir\left(\left(\sum_i \eta_{i0} + \gamma_0, \dots, \sum_i \eta_{i\ell} + \gamma_\ell \right)^T\right),
	\end{align*}
	where Dir(.) denotes the Dirichlet distribution \cite{Dirref}.
\item \label{item:gammasample} Sample each $\gamma_j, j = 0,\dots,\ell$, the parameters of the Dirichlet prior for $\vec\theta$, assuming an exponential hyperprior with parameter $s_j$, from the conditional posterior distribution:
	\begin{align*} \Pi(\gamma_j \mid \theta_j, \vec\gamma_{j^-}) \propto \frac{\Gamma(\sum_{k = 1}^\ell \gamma_k)}{\Gamma(\gamma_j)}
				\theta_j^{\gamma_j-1}\exp\{-s_j\gamma_j\},
	\end{align*}
	where $\vec\gamma_{j^-}$ denotes the vector or $\gamma$ parameters without the $j^{th}$ element.
\item \label{item:phisample} Sample each element in $\vec\phi_j = (\phi_{1j}, \dots, \phi_{jq})$, for all $\vec\phi_j, j = 1,\dots, \ell$.  If $\phi_{jk} \sim  g_{jk}(.; \vec\kappa_{jk}), k = 1, \dots, q$, the conditional distribution is as follows:
	\begin{align*}
		\Pi(\phi_{jk} \mid \mathbf W, \boldsymbol\eta, \vec\kappa_j) & \propto \prod_{i=1}^n p_{ij}(\vec W_i \mid \phi_{j}, \alpha, M = j)^{\eta_{ij}}g_{jk}(\phi_{jk}; \vec\kappa_{jk}),
	\end{align*}
	where $\vec\kappa_{jk}$ is the parameter vector for the prior distribution of $\phi_{jk}$.  
	
	In an exploratory examination of colonoscopy screening patterns observed in the SEER-Medicare data set, the hazard rate of the lag times is very flat (see Figure \ref{fig:flathaz}), so an exponential distribution for the lag times was used in our analysis of lifetime colonoscopy screening patterns (i.e. $f(t) = \lambda\exp\{-\lambda t\}$). In the exponential case, $\vec\phi_j = (\lambda_{j1}, \dots, \lambda_{jj})$, as each of the $j$ screenings has one associated parameter.  The cumulative hazard $\Lambda_{jk}(t) = \lambda_{jk} t$.  Let $\lambda_{jk} \sim Gamma(\kappa_{jk1}, \kappa_{jk2})$, with the following posterior distribution:
	\begin{align*}
		\Pi(\lambda_{jk} \mid \mathbf W, \boldsymbol\eta, \vec\kappa_{jk}) & \propto \prod_{i=1}^n p_{ij}(\vec W_i \mid \lambda_{jk}, \alpha, M = j)^{\eta_{ij}}
			\lambda_{jk}^{\kappa_{jk1}-1}\exp\{-\lambda_j/\kappa_{jk2}\},
	\end{align*}
	for $k = 1, \dots, j$ and $j = 1, \dots, \ell$.
\item \label{item:kappasample} Sample each element of $\vec\kappa_{jk}$, the prior parameters for $\phi_{jk}$ from the conditional posterior distributions.  In the exponential case, $\kappa_{jk} = (\kappa_{jk1}, \kappa_{jk2})$, with $\kappa_{jk1} \sim Exp(b_{jk})$ and $\kappa_{jk2}\sim IG(c_{jk}, d_{jk})$, where $IG(.)$ is the Inverse Gamma distribution. In this instance, the conditional posterior distributions become:
	\begin{align*}
		\Pi(\kappa_{jk1} \mid \lambda_{jk}, \kappa_{jk2}, b_{jk}) & \propto \left(\kappa_{jk2}^{\kappa_{jk1}} \Gamma(\kappa_{jk1})\right)^{-1} 
				\lambda_{jk}^{\kappa_{jk1}-1} \exp\{-b_{jk}\kappa_{jk1}\}\\
		\Pi(\kappa_{jk2} \mid \lambda_{jk}, \kappa_{jk1}, c_{jk}, d_{jk}) & \propto IG(\kappa_{jk1} + c_{jk}, \lambda_{jk}+ d_{jk})
	\end{align*}
\item \label{item:alphasample} Sample the correlation parameter, $\alpha$, from the conditional posterior
	\begin{align*}
		\Pi(\alpha \mid \mathbf W, \boldsymbol \eta,  \boldsymbol\phi, \vec\tau) & \propto \prod_{i=1}^n \left(p_{ij}(\vec W_i \mid \vec\phi_j, \alpha, M = j)  \right)^{\eta_{ij}}\times
			\alpha^{\tau_1-1}(1-\alpha)^{\tau_2-1},
	\end{align*}
where $\alpha \sim Beta(\tau_1, \tau_2)$.
\item \label{item:tausample} Sample the prior parameters for $\alpha$, $\tau_1$ and $\tau_2$, with both parameters distributed $Exp(1)$, such that the conditional posterior distributions are:
	\begin{align*}
		\Pi(\tau_1 \mid \alpha, \tau_2) &\propto \frac{\Gamma(\tau_1+\tau_2)}{\Gamma(\tau_1)} \alpha^{\tau_1-1}\exp\{-\tau_1\}\\
		\Pi(\tau_2 \mid \alpha, \tau_1) &\propto \frac{\Gamma(\tau_1+\tau_2)}{\Gamma(\tau_2)} (1-\alpha)^{\tau_2-1}\exp\{-\tau_2\}
	\end{align*}
\item \label{item:updateeta} Update each unobserved $\eta_{ij}, i = 1,\dots, n, j = 1,\dots, \ell$ with the expectation, using the Gibbs sampler estimates of the parameters at that iteration.  
\end{enumerate}

\subsection{Covariates}  \label{sec:gibbsestcovs}
Following previous work done by Ghitany and Maller \cite{GhitMalCov} and others \cite{GhitMalCov, IbrahimBook, ChenIbrPaper}, covariates are added to the model by incorporating them into the parameter(s) of interest.  The probability an individual $i$ has $j$ lifetime colonoscopy screenings can be modeled as $\theta_{ij} = \textrm{expit}(X_i' \vec\beta_j)$, where $X_i$ is a $1 \times p_j$ covariate vector for the $i^{th}$ subject, and $\vec \beta_j = (\beta_1, \dots, \beta_{p_j})$ are the effects of the $p_j$ covariates on the probability of $j$ lifetime colonoscopy screenings.  The expit(.) function, defined as 
$$ \textrm{expit}(a) = \frac{\exp\{a\}}{1+\exp\{a\}},$$
is used to ensure that the resulting $\theta_{ij}$ estimates will be between 0 and 1.

The elements of the parameter vector $\vec\phi_j$ can be modeled in a similar fashion using an appropriate link function.  If $\vec\phi_j$ only has one element, then $\phi_{ij} = h^{-1} \left(Z_i' \vec \omega_j\right),$ where $Z_i$ is the covariate vector for the $i^{th}$ subject, $\vec \omega_j$ are the effects of the covariates on the lag time, and $h(\cdot)$ is an appropriate link function.  (If $\vec\phi_j$ has more than one element, each element may be modeled with the same covariates and link function, or this may vary based on the constraints on the parameters in $\vec\phi_j$ and the biological rationale behind the covariate modeling.) In the exponential example, the link function $h(.)$ needs to be such that the parameter $\lambda_j$ is positive.  A natural function that ensures this is the exponential link function $\lambda_j = \exp\{Z_i' \vec \omega_j\}$, so that the resulting $\lambda_j$ are all greater than 0.

The likelihood function in equation \ref{eqn:multLik} can then be rewritten as follows:
\begin{align*}
	L(\mathbf W, \mathbf X, \mathbf Z, \boldsymbol \eta  \mid \boldsymbol \omega,\boldsymbol \beta, \alpha) & = \prod_{i=1}^n \prod_{j=0}^\ell \Bigg(\textrm{expit}\left(X_{i}' \vec\beta_j \right)
		p_{ij}\left(\vec W_i \mid h^{-1} \left(Z_{i}' \vec\omega_j\right), \alpha, M = j\right)\Bigg)^{\eta_{ij}} 
\end{align*}
A similar Gibbs sampler routine to that presented in \ref{sec:gibbsest} is used for estimation.  However, instead of sampling $\vec\theta$, each element of $\vec\beta_j, j = 1, \dots, \ell,$ is sampled from the posterior
\begin{align*}
\Pi(\beta_{jk} \mid \mathbf X, \boldsymbol \eta, \vec\beta_{j}^-, ) & = \prod_{i=1}^n  \left(\textrm{expit}\left(X_{i}' \vec \beta_j \right)\right)^{\eta_{ij}} 
			\frac{(\beta_{jk}-\mu_{\beta_{jk}})^2}{ \sigma^2_{\beta_{jk}}}, k = 1, \dots, p_j,
\end{align*}
with $\beta_{jk} \sim \mathcal{N}(\mu_{\beta_{jk}}, \sigma^2_{\beta_{jk}}), k = 1,\dots, p_j$.

Similar methods can be used for the estimation of $\omega_j, j = 1,\dots,\ell,$ replacing the steps for sampling $\phi_j$ and the prior parameters with steps for sampling each element of  $\omega_j$ for all $\omega_j$ and the associated prior parameters.

\section{Simulation Studies}
To determine the efficiency, accuracy and consistency of our method and algorithm in the SEER-Medicare data context, we conducted a simulation study for the multivariate screening case. We set the maximum number of possible lifetime screenings at two ($\ell = 2$), which is the maximum number of observed colonoscopy screenings in the SEER-Medicare data set, and is a value consistent with medical practice in the oldest old.  Data were generated varying the percentages of 0, 1, or 2 lifetime screenings, and assuming different lag times for subjects with only one screening when compared to subjects with two screenings.  As is suggested by the SEER-Medicare data (see Figure \ref{fig:flathaz}), we assumed an exponential distribution for the lag times.  Under the exponential distribution for the lag times, the multivariate survival distribution for subjects with two screenings becomes:
\begin{eqnarray}
	P(T_1 > t_1, T_2 > t_2 \mid \vec\lambda_2, \alpha, M = 2) = \exp\{-(\lambda_{21}t_1 + \lambda_{22}t_2)^\alpha\},\label{eqn:multSurvExp}
\end{eqnarray}
where $\vec\lambda_2 = (\lambda_{21}, \lambda_{22})$, and  $\lambda_{21}t_1$ is the cumulative hazard for the lag time to the first of two screenings, and $\lambda_{22}t_2$ is the cumulative hazard for the lag time to the second of two screenings.  For subjects who only receive one lifetime screening, the survival function reduces to the standard survival function for the exponential distribution, and is given by $$P(T > t \mid \lambda_1, M = 1) = \exp\{-\lambda_{1}t\}.$$  

To prevent identifiability issues, we assumed the maximum possible lag time was ten years (as subjects who are overdue by more than ten years are no longer ``average-risk" due to the rate of colorectal cancer progression \cite{WinawerCRCwork}), and that subjects were only eligible for colonoscopy screenings between the ages of 50 and 90 years old.  It is rare that a colonoscopy screening would be recommended for a patient over 90 because the risks associated with the colonoscopy screening procedure outweigh the long-term benefits of colonoscopy \cite{usptf1}.   The length of the simulated study was 25 years.  Three different lag time scenarios were used to generate data, and are denoted as `LT1', `LT2', and `LT3', (see Table \ref{table:LTS}). The three different lag time scenarios were paired with two different scenarios for the number of screenings (denoted as `NLS1' and `NLS2'), and can be seen in Table \ref{table:NLS}.  The correlation parameter $\alpha$ was set at 0.9 (light correlation between screenings), as was evidence by the SEER-Medicare data.  The parameter values used to simulate data were chosen based on observed lag times in the SEER-Medicare data set and the possible true number of lifetime screenings.  Paired together, there were six different types of simulated data sets, each containing 1000 subjects and generated 200 times. Left- and right-censoring percentages were approximately 50\% and 40\%, respectively, in the simulated data sets.  About 40\% of subjects had at least one observed screening, and  about 15\% of subjects had two observed screenings.  

Markov Monte Carlo (MCMC) chains were run on each data set, with the first 10,000 iterations burned for a total of 40,000 thinned iterations in each chain for analysis.  Point estimates were calculated as the median of the marginal posterior distribution of each parameter.

\begin{table} 
	\center
	\begin{tabular}{|c|c|c|c|}\\
		\hline
		\multirow{2}{*}{Scenario}&Only 1 Screening&\multicolumn{2}{|c|}{Two Screenings}\\
		\cline{2-4}
		&Lag Time ($\lambda_{11}$)&$1^{st}$ Lag time ($\lambda_{21}$)&$2^{nd}$ Lag time ($\lambda_{22}$)\\
		\hline
		LT1&4.3 (0.02)&1 (0.70)&1 (0.70)\\
		LT2&3.5 (0.09)&4/3 (0.50)&2/3 (1.05)\\
		LT3&2.25 (0.35)&4/3 (0.50)&2/3 (1.05)\\
		\hline
	\end{tabular}
	\caption{\footnotesize{Three different lag time (``LT") scenarios used to generate datasets used in our simulation study.  The lag times are shown are the median lag times (in years) from the survival distribution that is used to generate the lag time to colonoscopy screenings for individuals who are screened at least once in their lifetime.}}
	\label{table:LTS}
\end{table}
\begin{table} 
	\center
	\begin{tabular}{|c|c|c|c|}\\
		\hline
		\multirow{2}{*}{Scenario}&Probability of&Probability of&Probability of\\
		&0 Screenings ($\theta_0$)&1 Screening ($\theta_1$)&2 Screenings ($\theta_2$)\\
		\hline
		NLS1&1/3&1/3&1/3\\
		NLS2&0.5&0.25&0.25\\
		\hline
	\end{tabular}
	\caption{\footnotesize{Two different scenarios for the number of lifetime screenings (``NLS") used to generate datasets the number of lifetime screenings (0, 1, or 2) for each subject in the simulated data sets.}}
	\label{table:NLS}
\end{table}

\subsection{Simulation Results}

Performance of the algorithm was assessed through the bias and the square root of the mean square error (RMSE).  The bias was calculated as the average difference between the parameter estimate and the true value of the parameter, and the RMSE was calculated as the square root of the average squared difference between the parameter estimate and the true value of the parameter.  

Across all data set variations, the average bias (RMSE) of the median lag times, $\tilde{Y}_{11}, \tilde{Y}_{21}$ and $\tilde{Y}_{22}$, were 0.03 (0.41), 0.20 (0.70), and 0.91 (1.89) years for the only lifetime screening, the first of two screenings, and the second of two screenings.  Over a possible 10 year lag time, this represents an RMSE of 4.1\%, 7\%, and 18.9\% of the total possible range.  The bias and RMSE of  $\tilde{Y}_{11}$, the median of the lag time to the only lifetime screening, change little across the different LT and NLS models.  The median lag time with the largest RMSE is the second of two lag times, as it has the fewest observations available to contribute to parameter estimation, with the largest values occurring in the data sets with the longest lag times and lower probability of two observed screenings.  However, the overall RMSE is small, with less than a 1-year bias (over a 10-year range) across all data sets.

The bias (RMSE) for the percentage of screenings is small, with values equal to $-9.67\times10^{-4}$ (0.02), -0.01 (0.02), and 0.01 (0.03) for the probability of no lifetime screenings ($\theta_0$), the probability of one lifetime screening ($\theta_1$), and the probability of two lifetime screenings ($\theta_2$), respectively.  While $\theta_2$ has the largest RMSE, the overall quantity (3\%) is very small.

Table \ref{table:simulResults_bias} presents the bias (RMSE) of the parameter estimates across all simulated data sets.  This method precisely estimates the screening rates and the lag times, although larger bias and RMSE values are seen with the percentage of subjects with two screenings, and in the second lag time.  This is expected, as these parameters have the fewest screenings contributing observed data.

\begin{table} 
	\center
	\begin{tabular}{|c|l|cc|cc|cc|cc|}\\
		\hline
		\multicolumn{2}{|c|}{\multirow{2}{*}{Model}}&\multicolumn{2}{|c|}{LT1}&\multicolumn{2}{|c|}{LT2}&\multicolumn{2}{|c|}{LT3}&\multicolumn{2}{|c|}{Total (over $\theta$)}\\
		\cline{3-10}\cline{3-10}
		&&Bias&RMSE&Bias&RMSE&Bias&RMSE&Bias&RMSE\\
		\hline\hline
		\multirow{7}{*}{NLS1}&$\theta_0$&0.00&0.02&0.00&0.02&0.00&0.02&0.00&0.02\\
		&$\theta_1$&-0.01&0.03&0.00&0.02&-0.01&0.02&-0.01&0.02\\
		&$\theta_2$&0.02&0.03&0.00&0.02&0.01&0.02&0.01&0.03\\
		\cline{2-10}
		&$\tilde{Y}_{11}$&-0.02&0.59&-0.02&0.28&0.00&0.25&-0.02&0.42\\
		&$\tilde{Y}_{21}$&0.20&0.77&0.27&0.71&0.34&0.99&0.28&0.84\\
		&$\tilde{Y}_{22}$&1.02&2.00&0.32&0.90&0.54&1.43&0.66&1.55\\
		\cline{2-10}
		&$\alpha$&0.03&0.11&0.00&0.10&0.02&0.09&0.01&0.10\\
		\hline\hline
		\multirow{7}{*}{NLS2}&$\theta_0$&-0.01&0.02&0.00&0.02&0.00&0.02&-0.01&0.02\\
		&$\theta_1$&-0.01&0.03&0.00&0.02&-0.01&0.02&-0.01&0.02\\
		&$\theta_2$&0.03&0.04&0.00&0.02&0.01&0.03&0.01&0.03\\
		\cline{2-10}
		&$\tilde{Y}_{11}$&0.14&0.50&0.05&0.38&0.05&0.23&0.07&0.40\\
		&$\tilde{Y}_{21}$&0.05&0.11&0.18&0.69&0.09&0.45&0.12&0.52\\
		&$\tilde{Y}_{22}$&1.93&2.78&0.35&1.17&1.14&2.20&1.16&2.17\\
		\cline{2-10}
		&$\alpha$&0.06&0.09&0.02&0.07&0.04&0.07&0.04&0.08\\
		\hline\hline
		\multirow{7}{*}{Total (over the lag times)}&$\theta_0$&-0.01&0.02&0.00&0.02&0.00&0.02&0.00&0.02\\
		&$\theta_1$&-0.01&0.03&0.00&0.02&-0.01&0.02&-0.01&0.02\\
		&$\theta_2$&0.02&0.03&0.00&0.02&0.01&0.03&0.01&0.03\\
		\cline{2-10}
		&$\tilde{Y}_{11}$&0.04&0.57&0.02&0.34&0.03&0.24&0.03&0.41\\
		&$\tilde{Y}_{21}$&0.13&0.57&0.23&0.72&0.23&0.79&0.20&0.70\\
		&$\tilde{Y}_{22}$&1.51&2.45&0.37&1.09&0.86&1.88&0.91&1.89\\
		\cline{2-10}
		&$\alpha$&0.04&0.10&0.01&0.09&0.03&0.08&0.03&0.09\\
		\hline
	\end{tabular}
	\caption{\footnotesize{Summary bias and RMSE results for the simulation studies across all data sets.  Results of the simulations show the estimates of the screening rates to be close to the true parameter values, with RMSE values less than or equal to 3\% for all possible values of $\theta$, and ranging from 3 months to 2.5 years  for the median lag times, denoted as $\tilde{Y}$ (which is equal to 2.5\% to 20\% of the possible lag time range).  The parameter with the highest variation and bias is the second of two median lag time, $\tilde{Y}_{22}$ (as a function of $\lambda_{22}$),  which is expected as the percentage of subjects with two observed screenings is small. }}
	\label{table:simulResults_bias}
\end{table}

\section{Application to SEER-Medicare Data}
\begin{figure}
	\centerline{\includegraphics[width=2.5in]{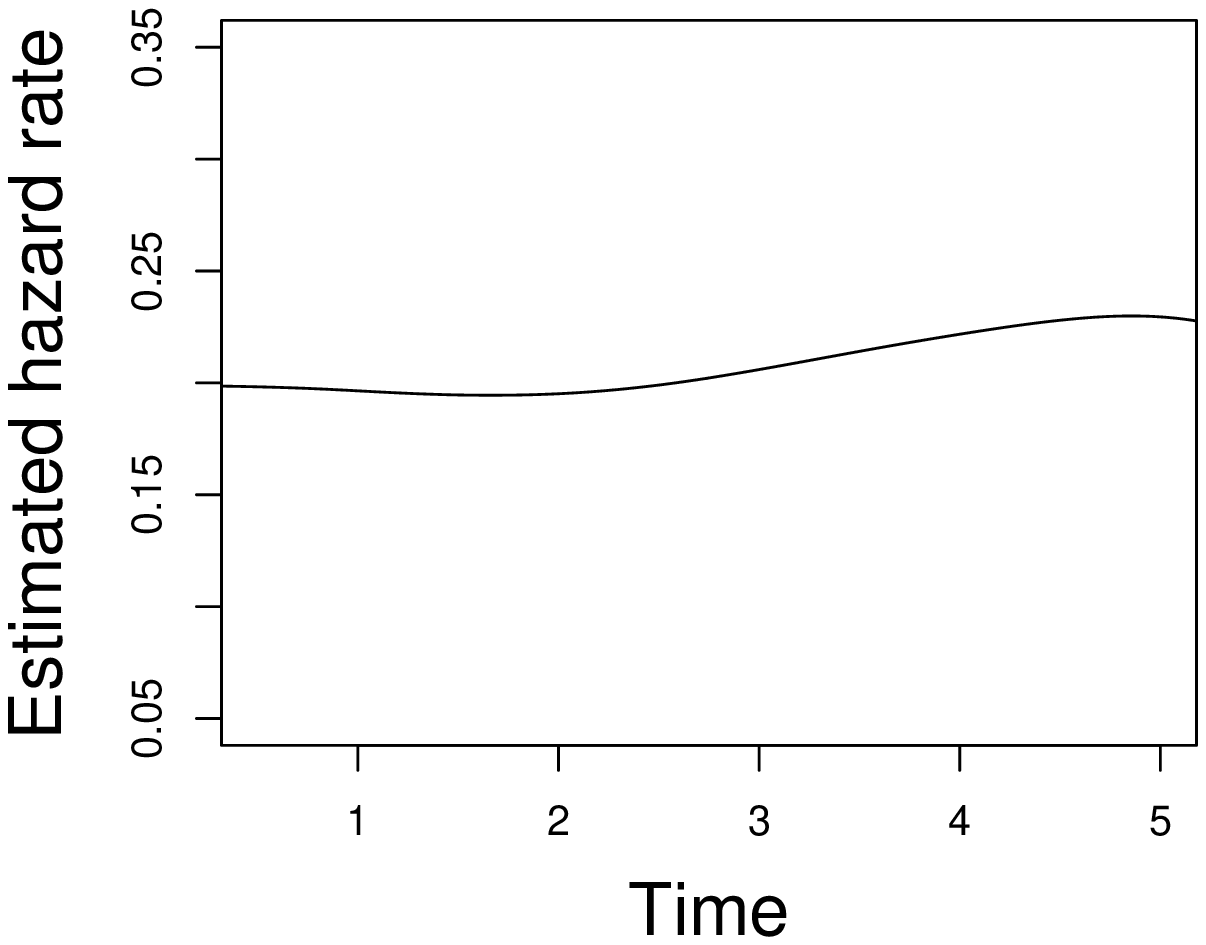}}
	\caption{\footnotesize{Approximated hazard rate of the time to the first screening between years 1991 and 1996 (before Medicare insurance coverage changes or guidelines were set).  The hazard rate is very flat, providing evidence that an exponential parametric distribution is appropriate.  The hazard rate is approximated dividing the number of observed failures by the number of subjects at risk (provided by \texttt{survfit()} in R), and then smoothed using \texttt{smooth.spline()} in R.}}
	\label{fig:flathaz}
\end{figure}

We applied our multivariate survival model  to the SEER-Medicare data set to investigate colonoscopy screening behavior between 1991 and 2003, assuming the maximum number of lifetime screenings was equal to 2 (i.e. $\ell = 2$), as that is the maximum number we observed in our data set.  This data set contains 403,842 individuals age 65 or older at study entry after the removal of subjects who used other CRC screening methods (such as fecal occult blood tests or sigmoidoscopy).  Individuals were considered eligible for screening colonoscopy in 1991; while current screening guidelines recommend screening starting at age 50, very few people received colonoscopies before the early 1990's (as the USPSTF did not even provide official guidelines until 1996 \cite{usptf1}), and therefore the probability of an unobserved screening on an average-risk individual occurring before 1991 is very small.  Among these subjects, 62\% were left-censored, with average left-censoring times equal to 4.5 years (range: 0.1 - 9.9 years) .  In addition, 22\% had one observed colonoscopy, and 0.11\% had two observed colonoscopies.  Among individuals who had at least one colonoscopy, the median lag time before the first observed screening was 5.7 years (range: 0.01 to 10 years), and among individuals with two colonoscopies, the median lag time before the second observed screening was 0.1 years (range: 0.01 to 2.7 years).  An approximated hazard rate showed the rate of screening was constant (see Figure \ref{fig:flathaz}), so we assumed an exponential distribution for all lag times.  

We examined both univariate models (estimating the time to the first screening), including a covariate to account for insurance coverage levels, as well as multivariate models estimating parameters for zero to two screenings.  The univariate models were created to provide initial estimates of the median time to the first screening (regardless of whether it was the only screening or the first of two), as well as the probability an individual would never be screened.  In addition, we were able to include an insurance level coverage covariate in the univariate model, which allowed us to quantify the effects of at least some insurance coverage on the probability of receiving at least one lifetime colonoscopy screening.  We then examined multivariate models with two possible lifetime screenings.  This allowed us to quantify differences between the lag time to the first lifetime screening and the lag time to the first of two lifetime screenings.  We were also able to examine if the lag time to the first screening was longer or shorter than the lag time to the second screening.  Among both the univariate and multivariate models, we compared the results when individuals were eligible for screening until age 75 or eligible for screening until age 80, which are commensurate with current screening guidelines \cite{usptf1}.  In the multivariate model, we assumed a maximum lag time of ten years to prevent issues with identifiability.  

Five separate Markov chain Monte Carlo (MCMC) chains were run for each model, each with a burn-in of 5,000, leaving a total of 15,000 thinned iterations in each chain for analysis.  Convergence was determined through the Geweke diagnostic \cite{geweke}, graphical diagnostics (such as trace plots and density plots), and Gelman-Rubin tests \cite{gelmanrubin, gelmanrubinadd}.  Point estimates were calculated as the median of the posterior marginal distributions for each parameter, and 95\% central credible intervals were used for inference. 

\subsection{Univariate model results}
We first examined univariate models, which provided us with initial estimates on the probability of never receiving a colonoscopy screening and the time to the first screening (based on the likelihood function in equation \ref{eqn:univLik}).  The simple univariate model shows that before age 75, approximately 38\% (95\% CI: 37.6\% - 37.9\%) of the Medicare population gets a colonoscopy screening, with a median lag time (calculated based on $\lambda$) equal to about 5.2 years (95\% CI: 5.16 - 5.20 years) . These numbers change slightly when the maximum screening age is raised to age 80, with slight increases in screening rates as well as increases in median lag times (see Table \ref{table:realRes_univcovs}).   
\begin{table} 
	\center
	\begin{tabular}{|c|c|cc|cc|}\\
		\hline
		Univariate&\multirow{2}{*}{Parameter}&\multicolumn{2}{|c|}{Cap at age 75}&\multicolumn{2}{|c|}{Cap at age 80}\\
		\cline{3-6}
		Model&&Estimate&95\% CI&Estimate&95\% CI\\
		\hline\hline
		\multirow{2}{*}{Simple Model}&$\theta$&0.622&(0.621, 0.624)&0.554&(0.552, 0.556)\\
		&Median time to screening&5.166&(5.135, 5.197)&6.533&(6.486, 6.576)\\
		\hline\hline	
		\multirow{3}{*}{Covariate Model}&$\theta$, no coverage&0.639&(0.637, 0.641)&0.570&(0.568, 0.572)\\
		&$\theta$, some coverage&0.508&(0.502, 0.513)&0.404&(0.398, 0.411)\\
		\cline{3-6}
		&Median time to screening&5.416&(5.381, 5.451)&6.896&(6.844, 6.945)\\
		\hline
	\end{tabular}
	\caption{\footnotesize{Univariate model results showing the median and 95\% credible intervals (calculated as the 2.5\% and 97.5\% of the MCMC chain for each parameter) of the posterior distributions for the probability of never being screened for colorectal cancer ($\theta$), and the median lag time (as a function of $\lambda$) to the first screening for colorectal cancer in the SEER-Medicare data set.}}
	\label{table:realRes_univcovs}
\end{table}

To determine the impact of changes in levels of insurance coverage for colonoscopy screenings (i.e. differences between coverage phase 0, phase 1, and phase 2) on colonoscopy screening rates, we included a covariate in the estimate of $\theta$ in the following manner:
\begin{align*} 
	\theta & = \textrm{expit}\Big( \beta_0 + \beta_1I \{ \textrm{study entry after 1998} \}\Big).
\end{align*}
In the covariate model, the baseline group (represented by $\beta_0$) were subjects who became eligible for screening when no colonoscopy coverage was offered, and $\beta_1$ represents the change in this probability when some or all coverage was available. (A covariate was not included in the parameter for the lag time, as there was not enough information to reliably run the Gibbs Sampler for this particular data set.) Results from the covariate model show that screening rates increased almost 15 percentage points for subjects age 75 and younger when at least some insurance coverage was offered, and increased almost 17 percentage points for subjects 80 and younger when at least some insurance coverage was offered.  These results show that providing at least some insurance coverage for colonoscopy screenings dramatically improves the rate of screening (see Table \ref{table:realRes_univcovs}).

\begin{figure}
	\centerline{\includegraphics[width=5in]{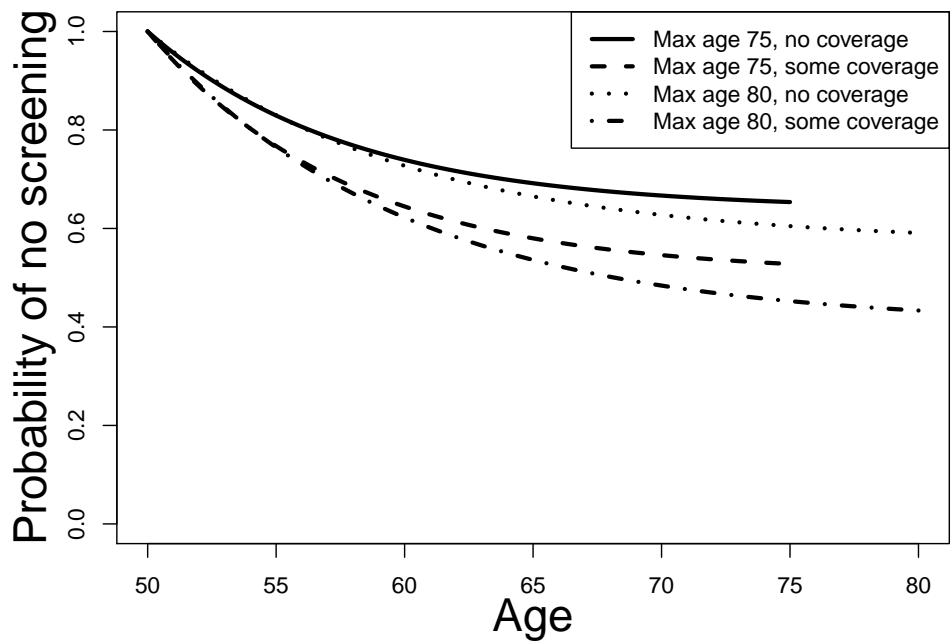}}
	\caption{\footnotesize{Univariate model results showing estimated survival curves (i.e. the probability of not being screened via colonoscopy) for the time to the first screening, comparing subjects with no colonoscopy insurance coverage to those with at least some colonoscopy insurance coverage.  In the colonoscopy screening context, a high survival curve indicates a poor rate of screening.  It can be observed that the subjects with the higher survival curve are those without colonoscopy coverage. }}
	\label{fig:StUnivCov}
\end{figure}

\begin{figure}
	\centerline{\includegraphics[width=5in]{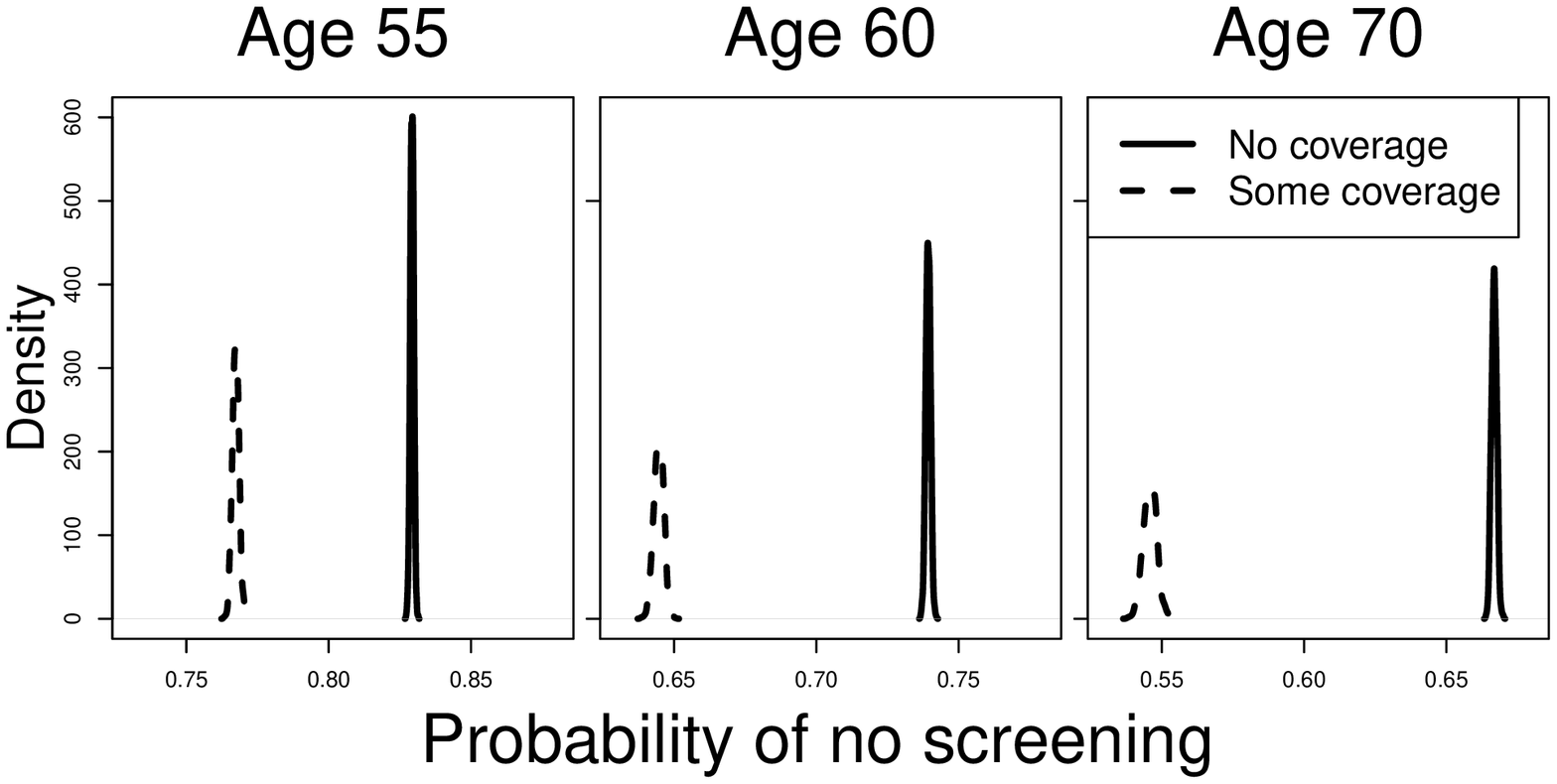}}
	\caption{\footnotesize{Univariate model results showing the densities for the probability of not being screened for colorectal cancer via colonoscopy by age 55, 60, and 70 years, comparing those with no coverage (solid line) and those with at least some coverage (dashed line).  Note that while the x-axes on all three graphs cover a different range of probability values, the size of the range is 15\% for all three graphs.  We can observe that the two densities are closest together at 55 years, and are farthest apart at 70 years.  Note that in all three figures, the densities do not overlap, providing evidence that the probability of screening under some and no coverage is statistically significantly different across the different time points.  The densities are calculated using the \texttt{density()} function in R on the MCMC chain of survival probabilities, calculated at each iteration of the thinned and burned chains.. Results are shown for the model with a maximum screening age of 75 years. }}
	\label{fig:StUnivCovDensity}
\end{figure}

Figure \ref{fig:StUnivCov} shows estimated ``survival curves" (i.e. the probability of no lifetime screening colonoscopy) for subjects who have no colonoscopy insurance coverage compared to subjects who have some or all colonoscopy insurance coverage.  In our analysis of colorectal cancer screening, a higher survival curve indicates a worse screening pattern (i.e. lower numbers and longer lag times), and it can be observed that (not surprisingly) the subjects who had no colonoscopy coverage had lower rates of screening.  Among patients eligible for screening up to age 75, 26\% of patients without coverage were screened by age 60, and 33\% of patients without coverage were screened by age 70.  However, when at least some coverage for colonoscopy was available, 36\% of patients were screened by age 60, and 45\% of patients were screened by age 70.  This can also be observed in Figure \ref{fig:StUnivCovDensity}, which shows the densities of the survival curves (i.e. probability of no colonoscopy screening) for subjects at age 55, age 60, and age 70.  The figure shows that at age 55, the two densities are the closest together, and each density is narrow.  However, by age 70, the densities are farther apart from each other, meaning that differences in screening patterns between subjects with and without insurance coverage become bigger with increasing age.  Note that in all three graphs, the densities do not overlap, providing evidence that the probability of never being screened via colonoscopy is statistically significantly different when subjects have some insurance coverage compared to those who have no insurance coverage.
\subsection{Multivariate model results}
In the multivariate case, we examined the case of two maximum possible lifetime screenings, as we had no individuals with three or more observed colonoscopies in our data set, and we assumed both lag times were distributed exponentially.  No covariates were included in the multivariate model; by the nature of the multiple screenings model, the lag times and screening percentages at different points in the study inherently include temporal changes such as insurance coverage levels.  

Results show that up to age 75, the probability of never being screened is approximately 68\% (95\% CI: 67.6\% - 67.9\%).  The probability of one lifetime screening is about 27\% (95\% CI: 26.8\% - 27.0\%), and the probability of two lifetime screenings is about 5\% (95\% CI: 5.3\% - 5.4\%) .  Among subjects who are only screened once, the estimated median lag time is 2.5 years (95\% CI: 2.53 - 2.55 years).  Among subjects who are screened twice, the median lag time for the first screening is 1 year (95\% CI: 1.01 -1.06 years), and the median lag time for the second screening is 1.6 years (95\% CI: 1.57 - 1.62 years) .  The parameter $\alpha$, which represents the correlation between screenings, is equal to 0.92 (95\% CI: 0.912 - 0.923) , which means the within-subject correlation between screenings is low.  Numbers changed little when subjects were eligible for screening up to age 80 (see Table \ref{table:realRes_multiv} for model results).  Note that the probability of never being screened through colonoscopy before age 75 is similar in the multivariate and univariate models.  Lag times between the univariate and multivariate models differ because of the unrestricted possible maximum lag time in the univariate models.

\begin{table} 
	\center
	\begin{tabular}{|c|cc|cc|}\\
		\hline
		\multirow{2}{*}{Covariate}&\multicolumn{2}{|c|}{Cap at age 75}&\multicolumn{2}{|c|}{Cap at age 80}\\
		\cline{2-5}
		&Estimate&95\% CI&Estimate&95\% CI\\
		\hline\hline	
		Probability of no screenings&0.677&(0.676, 0.679)&0.686&(0.685, 0.688)\\
		Probability of one screening&0.269&(0.268, 0.270)&0.224&(0.222, 0.225)\\
		Probability of two screenings&0.054&(0.053, 0.054)&0.090&(0.089, 0.091)\\
		\hline\hline
		Median time to only 1 screening&2.536&(2.526, 2.547)&2.687&(2.676, 2.698)\\
		Median time to $1^{st}$ of two screenings&1.037&(1.013, 1.064)&1.270&(1.253, 1.289)\\
		Median time to $2^{nd}$ of two screenings&1.597&(1.572, 1.620)&1.564&(1.540, 1.585)\\
		\hline\hline
		Correlation parameter $\alpha$&0.917&(0.912, 0.923)&0.961&(0.956, 0.964)\\
		\hline
	\end{tabular}
	\caption{\footnotesize{Median estimates and 95\% credible intervals (calculated as the 2.5\% and 97.5\% of the MCMC chain for each parameter) for the probability of receiving none, one, or two screenings in a lifetime, and the median time to the only screening or the first and second of two screenings for colorectal cancer in the SEER-Medicare data set, as well as the parameter $\alpha$, which represents the correlation between screenings.  Results are similar regardless of the maximum eligible age for screening.  }}
	\label{table:realRes_multiv}
\end{table}

\begin{figure}
	\center
	\includegraphics[width=6in]{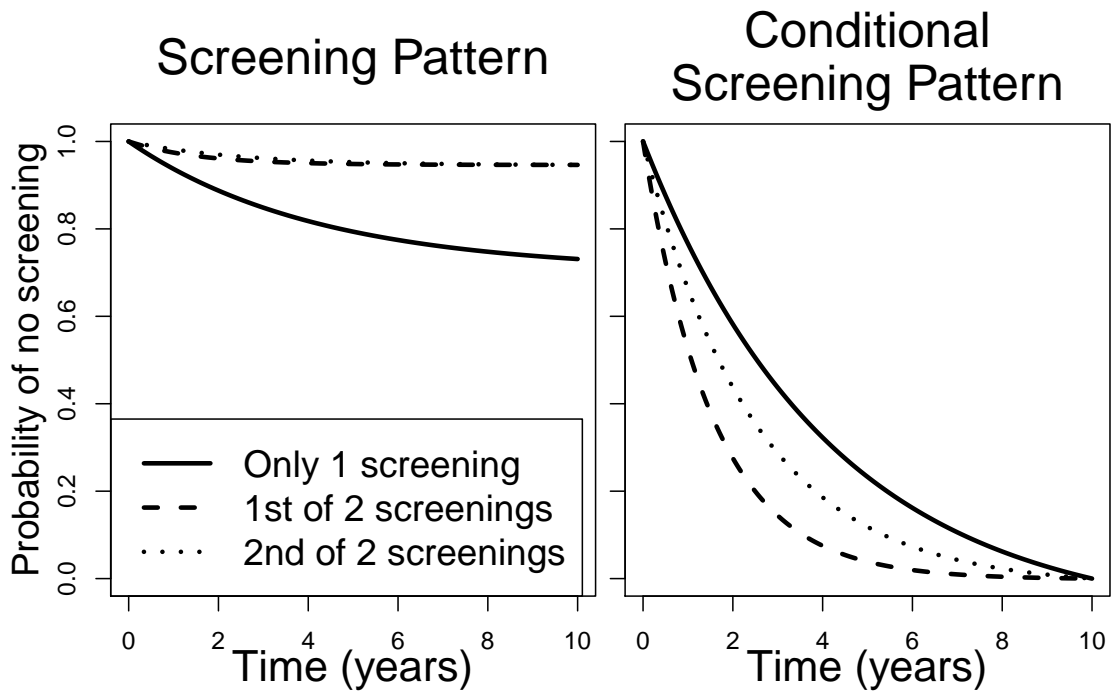}
	\caption{\footnotesize{LEFT: Multivariate model results showing the marginal estimated survival curves (i.e. probability of not being screened for colorectal cancer via colonoscopy) for the only lifetime screening (solid line), or the first and second of two screenings (dashed and dotted lines).  As expected, more people get only one screening in a lifetime rather than two lifetime screenings, and therefore that survival curve is lowest.  RIGHT: The estimated survival curves conditional on the number of lifetime screenings.  These curves show that among subjects who will receive two screenings, the first screening happens quickly when compared to the second screening.  Subjects who only get one colonoscopy take the longer than those who get two colonoscopies.   As with the univariate models, a high survival curve indicates poor screening rates.  Estimates were calculated from the multivariate model that assumes 75 is the maximum eligible age for screening.}}
	\label{fig:StMult}
\end{figure}

\begin{SCfigure}
	\includegraphics[width=3in, angle = -90]{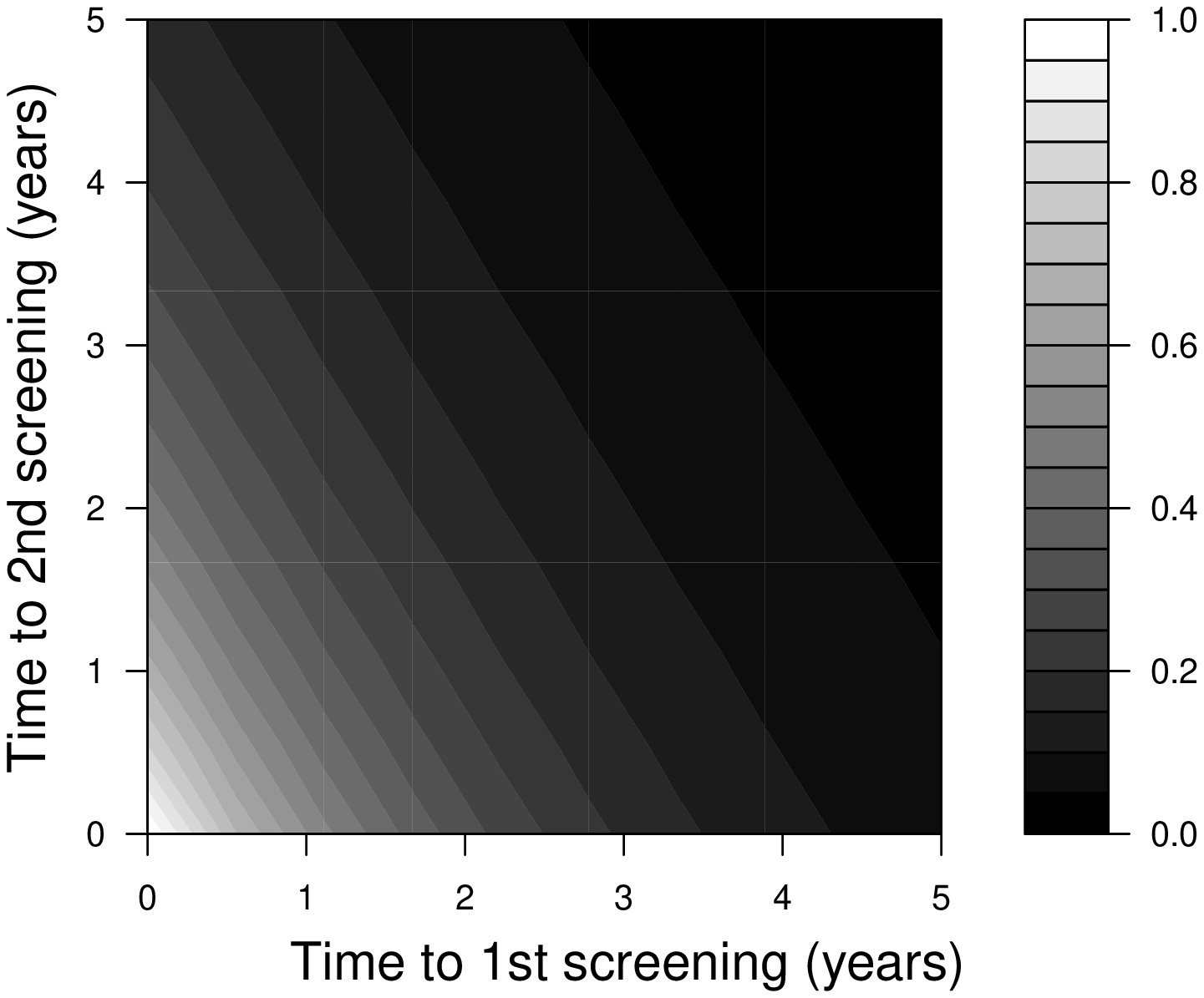}
	\caption{\footnotesize{A contour plot of the multivariate model results showing the joint survival curve (i.e. the probability of not being screened via colonoscopy), conditional on subjects who get two screenings (i.e. the probability of two screenings, $\theta_2$, is not used in the calculation of the survival probabilities) for the first five years.  The figure shows minor differences between the time to the first screening and time to the second screening, as the contour plot is fairly symmetric.  However, the grey shading extends slightly higher up the y-axis (which represents the time to the second screening), meaning that the probability of not being screened is higher for a longer time period before the second screening.  As with the univariate models, a high survival curve indicates poor screening rates.  Years 5 through 10 are omitted from the figure as the probabilities are very small difficult to discern in the figure.  Estimates were calculated from the multivariate model that assumes 75 is the maximum eligible age for screening.}}
	\label{fig:StMult_bivariate}
\end{SCfigure}

Estimated marginal survival curves (i.e. the probability of not being screened) for the each of the lag times of the multivariate model are shown in Figure \ref{fig:StMult}.  As with the univariate model, high survival curves indicate a poor screening rate.  On the left, it is observed that the number of two or more screenings is very low, and differences between the first and second of two screenings are minor.  The number of individuals receiving one lifetime colonoscopy screening is higher, but still poor.  Five years after becoming eligible for screening, 26\% of subjects have had one screening (either the only lifetime screening, or the first of two lifetime screenings).  Five years after becoming due for the second screening, only 4.7\% of subjects will have had a second screening.  On the right, the survival curves are again shown, but \textit{conditional} on getting one or two lifetime screenings (i.e. $\theta$ is not used in the calculation of the survival curve).  These estimates show that among individuals who get two screenings, the time to the first screening is shorter than the time to the second screening, with 96\% of these subjects getting the first screening within 5 years of becoming due, and 88\% of subjects getting the second screening within 5 years of becoming due.  Subjects who only get one lifetime colonoscopy screening waited longer, with 76\% of these subjects getting screened within the first five years of becoming due.   Our results show that while the actual rates of screening are poor, those who are getting screened are diligent, with a large majority of individuals getting screened within five years after becoming due.  Figure \ref{fig:StMult_bivariate} shows the bivariate survival distribution in a contour plot.  While the figure is fairly symmetric (meaning there is little difference between the lag time to the first screening and second screening), the grey shading extends slightly higher up the y-axis (the axis that denotes the time to the second screening), which means that the time to the second screening is delayed slightly longer when compared to the first screening.  The bivariate survival distribution is only shown for the first five years, as the probabilities for years five through ten are very small and it is difficult to discern differences in the distribution after this time point.

\section{Discussion}
We have proposed a cure rate model for multivariate survival data that can account for both left- and right-censored data. We have demonstrated theory that works for the general case of multiple lifetime screenings, and then applied it to the case of two colonoscopy screenings.  The case of two colonoscopy screenings in a lifetime is common, as beyond a certain age, the risks outweigh the long-term benefits of screenings, and are often not recommended in the later stages of life.  This model provides robust estimates, even in the difficult setting of considerable left- and right-censoring, and with the inclusion of subjects who never get screened.  Our approach provides estimates sufficiently accurate to detect both demographic differences and the time-varying impact of policy shifts.  

Using this method, we have shown that many individuals are never being screened for colorectal cancer, with overall estimates of at least one screening at only 30\%.  However, screening behavior was dramatically improved following increases in Medicare payments, with an estimated reduction in the probability of never being screened for colorectal cancer of around 15\% or more when colonoscopy coverage was provided.  These results agree with previous work, which has shown that screening incidence is generally low, but can be improved with increased levels of coverage \cite{HealthcareSystemFactors,ScrnDecisionMaking}.  In addition, among subjects who do get screened, they are diligent, and do not wait long periods of time after becoming due for a screening.  We have extended these results to quantify the exact rates of incidence and how adherent individuals are to current screening guidelines.

Future work with this model and the SEER-Medicare data set includes linking lifetime screening behavior to cancer incidence rates, as well as the inclusion of other screening modalities, such as sigmoidoscopy and FOBT.  This link will greatly inform the debate on optimal screening guidelines, as well as improve current cost-benefit analyses of CRC screening and Medicare expenditures.

We have only presented simulations for the case of two lifetime screenings, which is reasonable for analysis of the SEER-Medicare data set. The extension to 3 or more lifetime screenings is more difficult computationally, although it can be done with time and care.  Our model has answered very important questions about colorectal cancer screening behavior, but also has broad applicability to situations with multiple events where there may be patterns unobserved before study entry or after study exit.  These types of analyses will become more prevalent as time progresses, particularly with major changes in health care coverage due to the Affordable Care Act.  Accurate assessment of patterns of lifetime preventive medical care will become more necessary as government-funded health care becomes more prevalent, and this information is required to determine the effectiveness of different medical procedures.  

\section*{Acknowledgements}
Dr. Beckett acknowledges support from the UC Davis Cancer Center Support Grant, P30CA093373-06.  Statistical support was made possible by Grant Number UL1 RR024146 from the National Center for Research Resources (NCRR), a component of the National Institutes of Health (NIH), and NIH Roadmap for Medical Research. Its contents are solely the responsibility of the authors and do not necessarily represent the official view of NCRR or NIH.  Information on Re-engineering the Clinical Research Enterprise can be obtained from http://nihroadmap.nih.gov/clinicalresearch/overview-translational.asp.  

The authors would like to thank Dr. Joshua Fenton at the UC Davis Medical Center for his guidance and support with this research.

\end{document}